\documentclass[aps,pra,amsmath,twocolumn,showpacs,amssymb,superscriptaddress]{revtex4}
\usepackage{times}
\usepackage{amsmath}
\usepackage{listings}
\usepackage{float}
\usepackage{amsfonts} 
\usepackage{graphicx}
\usepackage{afterpage}
\usepackage{amsthm}
\usepackage{todonotes}
\usepackage{color}
\usepackage{times}
\usepackage{bm}

\begin{document}
\title{Emergence of chaotic cluster synchronization in heterogeneous networks} 

\author{Rodrigo M. Corder}
\affiliation{Division of Epidemiology and Biostatistics, School of Public Health,University of California, Berkeley, Berkeley 94720, USA}

\author{Zheng Bian} 
\affiliation{Instituto de Ci\^encias Matem\'aticas e de Computa\c{c}\~ao, Universidade de S\~ao Paulo, S\~ao Carlos 13566-590, Brazil}
\thanks{Corresponding author is the third one:  tiago@icmc.usp.br}

\author{Tiago Pereira}
\affiliation{Instituto de Ci\^encias Matem\'aticas e de Computa\c{c}\~ao, Universidade de S\~ao Paulo, S\~ao Carlos 13566-590, Brazil}
\affiliation{Department of Mathematics, Imperial College London, London SW7 2AZ, UK}

\author{Antonio Montalb\'an}
\affiliation{Department of Mathematics, University of California, Berkeley, Berkeley 94720, USA} 

\date{\today}

\begin{abstract}
Many real-world complex systems rely on cluster synchronization to function properly. A cluster of nodes exhibits synchronous behavior while others behave erratically.  Predicting the emergence of these clusters and understanding the mechanism behind their structure and variation in response to parameter change is a daunting task in networks that lack symmetry. We unravel the mechanism for the emergence of cluster synchronization in heterogeneous random networks. We develop a heterogeneous mean field approximation together with a self-consistent theory to determine the onset and stability of the cluster.  Our analysis shows that cluster synchronization occurs in a wide variety of heterogeneous networks, node dynamics, and coupling functions. The results could lead to a new understanding of the dynamical behavior of networks ranging from neural to social.
\end{abstract}

\maketitle

{\bf   Synchronization is an important phenomenon in networks impacting communications, biology, chemistry, and physics. Extensive studies have addressed the onset of global synchronization and its relation to the interaction structure of the network and the node dynamics. Recent work reveals that cluster synchronization, where network interactions drive certain units to behave in unison while others exhibit erratic patterns, promotes health and coherence in real-world scenarios. While symmetry-induced cluster synchronization is characterized, its onset for networks that lack symmetry remains elusive. Our work unveils the phenomenon of chaotic units achieving sustained and stable cluster synchronization within heterogeneous networks. The initiation by hubs, followed by their desynchronization, leads to a stable cluster of moderately connected nodes. As coupling strengthens, nodes join or depart the cluster according to their connectivity degree. We introduce a novel heterogeneous mean-field approach and a self-consistent theory predicting cluster membership, stability, and formation mechanisms.}

Synchronization in complex networks is key for the proper functioning of  various real-world complex systems {\color{black} ranging from communication \cite{yadav2017self}, via biology \cite{winfree2001geometry} to chemistry \cite{sebek2016complex,kuramoto2003chemical}}.  Nodes of the network adapt their dynamical behavior because of the network interaction to move in unison.  While \textit{global} synchronization, where all units of the system behave in unison, has been deeply studied \cite{Eroglu_2017_sync_chaos,arenas2008,flunkert2010,omel2012,tonjes2021}, this behavior is often related to pathologies such as Parkinson \cite{hammond2007pathological} and epilepsy \cite{lehnertz2009synchronization}. In fact, most real-world systems rely on \textit{cluster} synchronization for their functioning. In this case, some units exhibit synchronous behavior while others behave erratically. Examples include multi-robot systems carrying out parallel tasks \cite{dey2022synchrosim} or neural systems where cluster synchronization is associated with the healthy state of the individual \cite{schnitzler2005normal}.

When such cluster synchronization results from a graph symmetry, recent progress allows one to determine the onset, membership, and stability of the clusters of synchronized nodes \cite{pecora2014cluster,siddique2018symmetry, 2303.08668}. However, synchronized clusters are prominent in networks such as neuron networks with connectivity structures that lack symmetries \cite{bonifazi2009gabaergic,pereira2010hub}.  Indeed, certain phase models on random heterogeneous networks exhibit a degree-dependent cluster formation where hubs serve as an engine for cluster formation \cite{mcgraw2005}. As the coupling increases, other nodes join the synchronized cluster, leading to a giant synchronized cluster  approaching global synchrony \cite{gomez2007paths,lee2005synchronization}. Interestingly, for general models in heterogeneous networks, global synchronization is unstable \cite{Matteo}. That is, for large coupling strengths, hubs lose their synchrony, and other nodes can display synchronization by forming their own cluster. All this can happen while the network behaves erratically, far from any global synchronization. Surprisingly, such cluster formation in random networks remains undisclosed.

Here, we uncover how chaotic units coupled on heterogeneous networks display sustained and stable cluster synchronization.  While in synchronous motion, the cluster remains enslaved by the chaotic dynamics of the mean field.  As the coupling strength increases, nodes can join and leave the cluster depending on their degree. We develop a heterogeneous mean field approach and a self-consistent theory capable of predicting which nodes belong to the cluster and its stability and shed light on the cluster formation mechanisms.

{\it Dynamics in a heterogeneous random network.}
Consider an undirected network $G$ on $N$ nodes defined by the adjacency matrix $A=\{A _{pq}\}$, where
$A _{pq}=1$ if nodes $p$ and $q$ are connected and $A _{pq}=0$ otherwise.
Each node $p$ supports isolated dynamics $f(z) = 2z\mod 1.$ The state $z_p^t$ of node $p$ at time $t$ evolves by the discrete-time Kuramoto model:
\begin{eqnarray}
z_p^{t+1} =
  f(z_p^{t}) + \frac{\alpha}{C} \sum_{q=1}^{N} A_{pq} \sin[2\pi\left( z_q^{t} - z_p^{t} \right)] \mod 1,
 \label{eq:Kuramoto}
\end{eqnarray}
where $\alpha$ is the coupling strength, and the network mean-degree $C$ normalizes the total interaction. Throughout the text, we  focus on this particular case of the isolated dynamics $f$ that is chaotic and stochastically stable \cite{Viana}. This means that under a small coupling, no cluster can be formed.  In Appendix \ref{sec_app:generalized}, we discuss the general isolated dynamics $f$. 

{\it Heterogeneous random networks.}
We construct the network $G$ from Chung-Lu \cite{CL06} random graph model $G(w)$ with expected degree sequence $w=(w_1,\cdots,w_N)$.
For $p>q$, each $A _{pq}$ is a Bernoulli variable  mutually independent with success probability $w_pw_q/\sum_{k=1}^N w_k$. It can be shown that when $N$ is large, the actual degree sequence of $G$ is concentrated around $w$. 
 Concretely in our numerical experiments, we prescribe $w$ with $N=5\times 10^5$ to follow an inverse Gamma distribution $\mathrm{Inv}\Gamma(2,C)$. In our realization, the mean degree is $C=300$, and the maximal degree of the hub is $11977$. The tail of the distribution is a power law with exponent $3$. 

{\it Order parameter.} To study the cluster formation, we introduce the ensemble order parameter taking into account the different roles played by each node in the network 
\begin{equation}\label{eq:Kura}
r  e^{i 2\pi \theta}:= \frac{1}{\mathrm{Vol}(G)}\sum_{p=1}^N\sum_{q=1}^N  A _{pq} e^{i 2\pi z_q},
\end{equation}
where
$\mathrm{Vol}(S):= \sum_{q\in S} w_q$ denotes the volume of a subgraph in $G$; in particular, $\mathrm{Vol}(G)=CN.$ Since the network is heterogeneous, the ensemble order parameter is suitable as it weighs the contribution of the nodes according to their degrees. When $r$ equals $1$, the whole network is perfectly synchronized. In heterogeneous networks, global synchronization is unstable \cite{Matteo}. {\color{black} Thus, cluster synchronization is the only possible, stable collective dynamics that provide a nonzero value for the amplitude $r$ of the order parameter.   }

\begin{figure}
	\includegraphics[width=85mm]{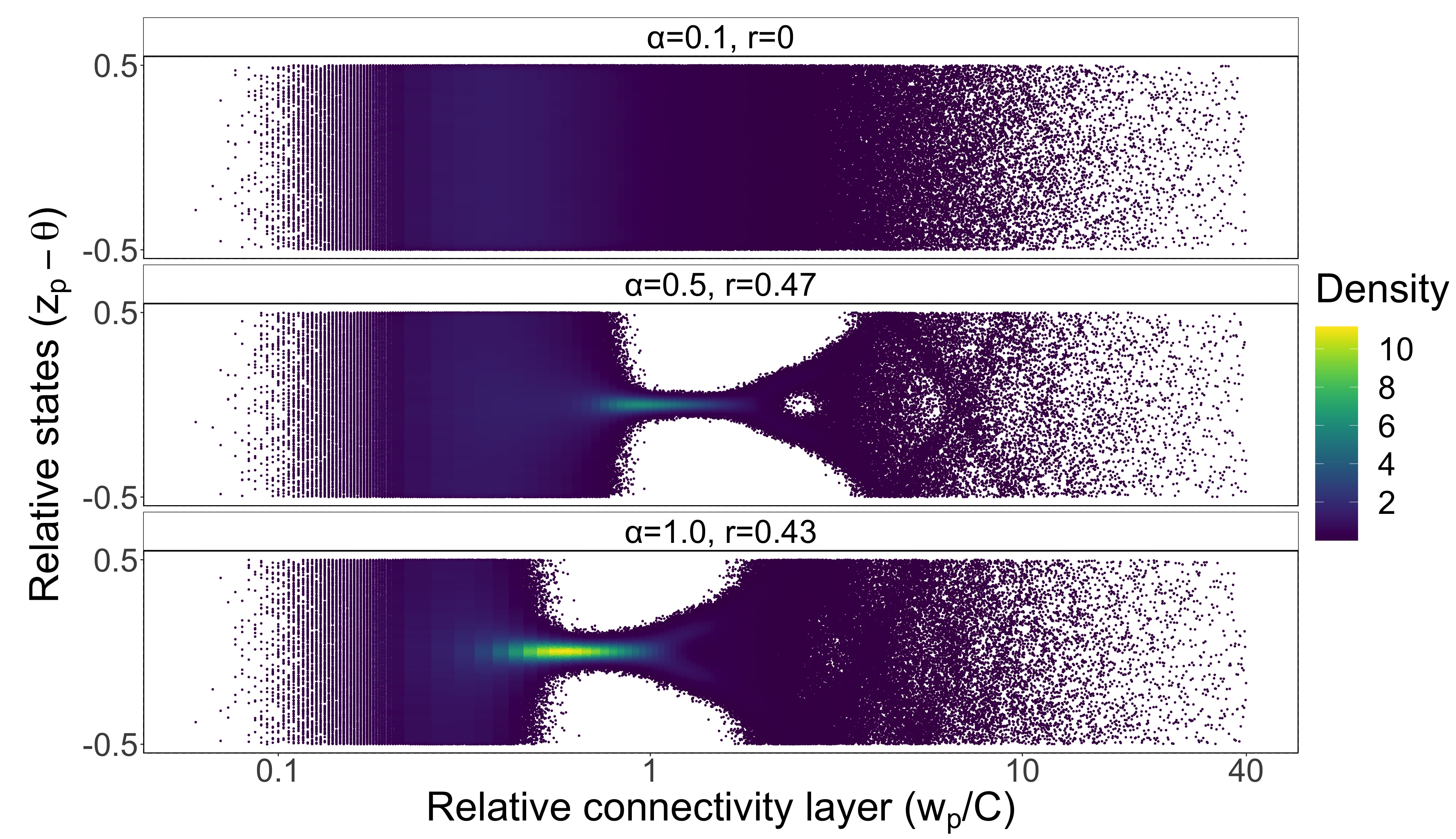}
	\caption{\label{fig:1} {\bf Emergence of spontaneous cluster synchrony}. We show three snapshots of the relative states $(z_p-\theta)$ at coupling strengths $\alpha=0.1$, $0.5$, and $1.0$ for a fixed network realization. The vertical axis represents the relative position on the circle $z_p-\theta$ with respect to global field phase $\theta$, and the horizontal axis the relative connectivity layer $w_p/C$. The bright colors emphasize synchrony to $\theta$. For weak coupling strength $\alpha=0.1$, the network dynamics do not admit cluster synchrony. As the coupling strength increases, cluster synchrony emerges at $\alpha=0.5$; furthermore, at $\alpha=1.0$, the cluster transforms as new nodes join while others leave.}
\end{figure}

{\it Cluster formation.} {Starting from uniformly distributed initial conditions,
we probe couplings $\alpha$ between $0$ and $1.2$ by iterating the network dynamics until reaching the stationary configuration of cluster synchrony, where the ensemble amplitude $r$ becomes stationary while the phase $\theta$ evolves in time.
Fig. \ref{fig:1} presents three snapshots of the stationary configuration of cluster synchrony at $\alpha=0.1, 0.5$ and $1.0$.} We plot in the horizontal axis the relative connectivity layer $w_p/C$, that is, gathering all nodes $p$ with the same relative degree $w_p/C$, and in the vertical axis the relative state  $z_p - \theta$. For $\alpha=0.1$, we observe a nearly uniform distribution. At $\alpha=0.5$, a group of nodes behaves in unison, synchronizing with $\theta$. The bright colors from the heat map indicate the concentration of points.
This behavior persists for large values of $\alpha$ such as $\alpha =1$.

{As the $\alpha$ increases,  high degree nodes desynchronize, and other nodes with smaller degrees join the cluster. In Fig. \ref{fig:1}, the bright colors shift towards the lower connectivity layers as $\alpha$ increases from $0.5$ to $1$. Other layers tend to follow the cluster but not too sharply, as can be observed in Fig. \ref{fig:1} by the spreading of the states in particular layers above the cluster. For a more detailed discussion of the other layers, see {  Appendix \ref{sec_app:sync_map}.}}

\begin{figure}
	\includegraphics[width=85mm]{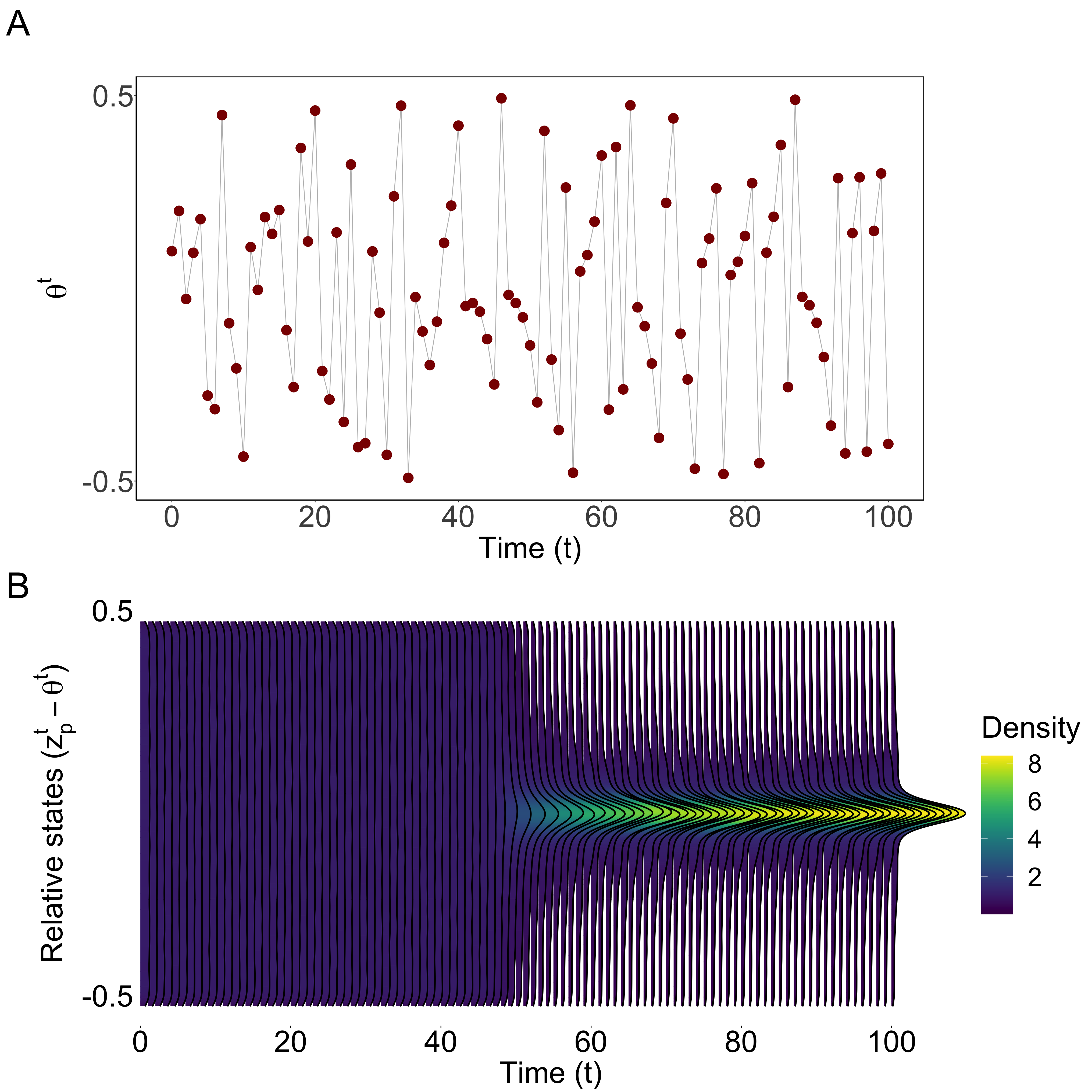}
	\caption{\label{fig:2} {\bf Time-evolution of the spontaneous emergence of cluster synchrony}. In panel A at $\alpha=1$, the time series of the mean field phase $\theta$ reveals a chaotic dynamics.  In panel B, for the same $\alpha$,
the histograms show the time-evolution of the relative states $(z_p^t-\theta^t)$ of nodes in the cluster $S_{\theta}$ across time $t$. Concentration near zero indicates that nodes in the cluster spontaneously synchronize with $\theta$. }
\end{figure}

{\it Cluster synchronization driven by mean field phase $\theta$}.
For coupling strengths $\alpha$ that admit cluster synchrony, { Fig. \ref{fig:1} suggests that the cluster dynamics synchronize to $\theta$, which, interestingly, has an erratic behavior as shown in Fig. \ref{fig:2}A.} {\color{black} To see the time-evolution towards this synchronization, we restrict the analysis to this cluster of nodes and denote it by $S_{\theta}$. For fixed $\alpha=1$,} starting from a uniformly distributed initial network state, we plot the histogram of the states for the nodes in the synchrony cluster $S_{\theta}$. Fig. \ref{fig:2}B shows a vertical histogram with increasing concentration near zero, indicated by thickness and bright color. Nodes in the cluster spontaneously come into synchrony towards $\theta$. 

{\it A heterogeneous mean-field approach}.
To analyze these findings we develop a theoretical approach capable of explaining the cluster formation and the enslavement of the cluster dynamics to a chaotic motion.
Informed by the stationary cluster synchrony configuration, we use the ansatz that \textit{there is a sustained cluster $S_{\theta}$ synchronizing to the global field phase $\theta$ at coupling $\alpha$, while the other nodes spread uniformly.} 
{It remains to determine which nodes belong to $S_{\theta}$, establish its stability, and analyze the dynamics of the mean field phase $\theta$. Already from the ansatz, we claim that the order parameter amplitude $r$ is stationary. Indeed, since $re^{i2\pi\theta}= (\sum_{q=1}^N d_q e^{i2\pi z_q})/\mathrm{Vol}(G)$, where $d_q$ is the actual degree of node $q$, and nodes that do not belong to $S_{\theta}$ provide a negligible contribution to the ensemble order parameter we obtain $r= (\sum_{q\in S_{\theta}} d_q)/\mathrm{Vol}(G)$. Here, we used that nodes in $S_{\theta}$ satisfy $z_p = \theta$. By concentration properties of the network, the actual degrees $d_q$ are asymptotically almost surely approximated by the ensemble average $\mathbb{E}d_q=w_q$, therefore
\begin{equation}\label{eq:r}
r = \frac{\mathrm{Vol}(S_{\theta})}{\mathrm{Vol}(G)}.
\end{equation}}
{\color{black} To determine which nodes belong to $S_{\theta}$, our first step is to write the network equations in terms of the ensemble order parameter.} We define the local field at each node $p$ to be $$r_pe^{i2\pi\theta_p}:= \sum_{q=1}^N A _{pq} e^{i2\pi z_q}.$$
Multiplying both sides by $ e^{-i 2\pi z_p}$ and comparing the imaginary parts, we can write the network dynamics as
\begin{equation}\label{LocalMean}
	z_p^{t+1} = f(z_p^t) + \frac{\alpha}{C} r_p^t \sin [2\pi(\theta_p^t - z_p^t)].
\end{equation}

{\color{black} The cluster synchrony ansatz implies that 
$\theta_p=\theta,$ and $r_p= \sum_{q\in S_{\theta}}A_{pq} $ equals the number of neighbors of node $p$ that belong to the cluster $S_{\theta}$. Again, by the network concentration properties, this number can be approximated by its ensemble average 
$\sum_{q\in S_{\theta}} \mathbb{E}A_{pq} = \sum_{q\in S_{\theta}}\frac{w_pw_q}{\mathrm{Vol}(G)} = w_p \frac{\mathrm{Vol}(S_{\theta})}{\mathrm{Vol}(G)},$ and
we obtain the heterogeneous mean field approximation:
\[
r_p = w_p r.
\]
}
 Notice that such approximation relies solely on the cluster synchrony ansatz and the concentration properties of our random network.
Plugging it into Eq. (\ref{LocalMean}) yields
\begin{equation}\label{hmf}
z_p^{t+1} = f(z_p^t) + \frac{\alpha}{C} w_p  r^t \sin [2\pi(\theta^t- z_p^t)].
\end{equation}
{ To obtain the $\theta$-dynamics, consider any cluster node $p \in S_{\theta}$ so that $z_{p} = \theta$ and hence the interaction term in Eq. (\ref{hmf}) vanishes, resulting in that $z_{p}$ and hence $\theta$ evolve by the isolated map. This explains the chaotic behavior of the mean field phase $\theta$. Next, we determine the stability of $S_{\theta}$ and estimate its size via a self-consistent theory. 

{\it Stability of the synchronous motion of the cluster $S_{\theta}$.} We established the stationarity of ensemble amplitude $r$. For now, let us assume its value $r>0$ is known and determine the cluster $S_{\theta}$ in terms of $r$ by studying the displacement $s_q^t = z_q^{t} - \theta^{t}$. Because of the particular form of $f$ and the heterogeneous mean field approximation we obtain 
\[
s_q^{t+1} = f(s_q^t) - \frac{\alpha }{C}w_q r\sin(2\pi s_q).
\]
The above map can be parametrized as $f_{\beta}(s) = f(s) - \beta \sin (2\pi s)$
and inserting the appropriate value of $\beta$ we recover the map for $s_q$.
{Observe that $s=0$ is always a fixed point of $f_{\beta}$ for any parameter $\beta$. To check its attractivity, we compute the linearization of  $f_{\beta}$  and obtain $f^{\prime}_{\beta}(0)=2- \beta 2\pi.$ Thus, the fixed point $s=0$ attracts when $\beta\in (1/2\pi,3/2\pi)$. We therefore obtain the condition for synchrony cluster 
$\frac{\alpha }{C}w_q r \in (1/2\pi,3/2\pi)
$, 
that is,
$z_q$ synchronizes to $\theta$ whenever
\[
w_q\in\left(\frac{C}{\alpha r 2\pi}, \frac{3C}{\alpha r 2\pi}\right).
\]
\noindent
This determines $S_{\theta}$ in terms of the amplitude of the ensemble order parameter $r$. For a bifurcation analysis of $f_{\beta}$ and its relation to the finer cluster structures, see {  Appendix \ref{sec_app:sync_map}}. }

{\it Self-consistent equation for $r$.} 
{Now we turn to the ensemble order parameter amplitude $r$, {which is key to predicting the nodes that belong to the cluster $S_{\theta}$}. We aim to determine the value of $r$ as a function of coupling strength $\alpha$, and will be particularly interested in its bifurcation from zero to positive values. Using Eq. (\ref{eq:r}), we can write  
$r =  \frac{1}{C}\int_{S_{\theta}} w\delta(w)\mathrm{d}w$,} where $\delta(w)$ is the probability density function for the degree distribution of the network $G$.

We notice that $S_{\theta}$ depends on $r$ and the values of $r$ must satisfy a self-consistent relation. Consider 
\begin{equation}\label{eq:R(r)}
r =R_{\alpha}(r):=\frac{1}{C}\int_{C/\alpha r 2\pi}^{3C/\alpha r 2\pi} w\cdot\delta(w)\mathrm{d}w.
\end{equation}
In fact, $R_{\alpha}(r)$ defined above is a first approximation of $r$ only by the cluster $S_{\theta}$ contribution. Further approximation can be constructed by considering nodes locking phase with $S_{\theta}$ and layers that are not uniformly distributed.
For more detailed derivation and discussion of the self-consistent equation, see Appendix \ref{sec_app:self-consistent}. We will determine $r$ by finding a fixed point of the map $R_{\alpha}$. In our case, the degrees follow an inverse gamma distribution $\mathrm{Inv}\Gamma(2,C)$, and evaluating the integral we obtain
\begin{align*}
R_{\alpha}(r) =  e^{-2\pi \alpha r}[e^{(4/3)\pi \alpha r} -1].
\end{align*}

Note $r_0=0$ is always a fixed point of $R_{\alpha}(r)$ for any $\alpha$ and
$R^{\prime}_{\alpha}(0) =\frac{4\pi\alpha}{3}$. Through a bifurcation analysis for $R_{\alpha}$, see {  Appendix \ref{sec_app:self-consistent}} for details, we identify three parameter regimes. (i) For $\alpha\in(0,\frac{3}{4\pi})$, $r_0=0$ is attractive  and no cluster synchrony. (ii) At $\alpha= \frac{3}{4\pi}$, the fixed point $r_0=0$ loses stability and gives rise to a new attractive fixed point $r>0$. Cluster synchrony emerges  among the layers of degree between $(C/2\pi \alpha r,3C/2\pi \alpha r)$.  (iii) As $\alpha$ increases beyond a threshold $\alpha_*\approx 2.1$, $r $ loses stability and bifurcates into an attractive period-2 orbit and further through a period-doubling cascade.

To pinpoint the emergence of cluster synchrony by the bifurcation into $r >0$, in Fig. 3 the solid line is the theoretically predicted $r $ found as the attractive fixed point of $R_{\alpha}$. The actual formula used is slightly different from Eq. (\ref{eq:R(r)}) to compensate for  {  the discrepancy due to the simplifying assumption that nodes outside the cluster are distributed uniformly}; for more details, see {  Appendix \ref{sec_app:self-consistent}}. The dots are the empirically calculated $r $ obtained by simulating the large heterogeneous network dynamics on $G$. We probe each $\alpha$ by forcing the network's initial condition into synchrony; for more details, see {  Appendix \ref{sec_app:forced_sync}}. We discard the first 2000 iterates as transient and collect the network states for the next 2000 iterates to compute the empirical $r$ according to Eq. (\ref{eq:Kura}) at each iterate and finally output the average value. Stationarity is confirmed by small standard deviations $< 0.025$ around the average. 

As mentioned earlier, we presented the case of the Bernoulli map $f$ as isolated dynamics and the Kuramoto coupling for the sake of simplicity. Both the heterogeneous mean field and the self-consistent approach generalize to further isolated dynamics, coupling functions, and degree distributions. We provide these details in the {  Appendix \ref{sec_app:generalized}}.

\begin{figure}
	\includegraphics[width=85mm]{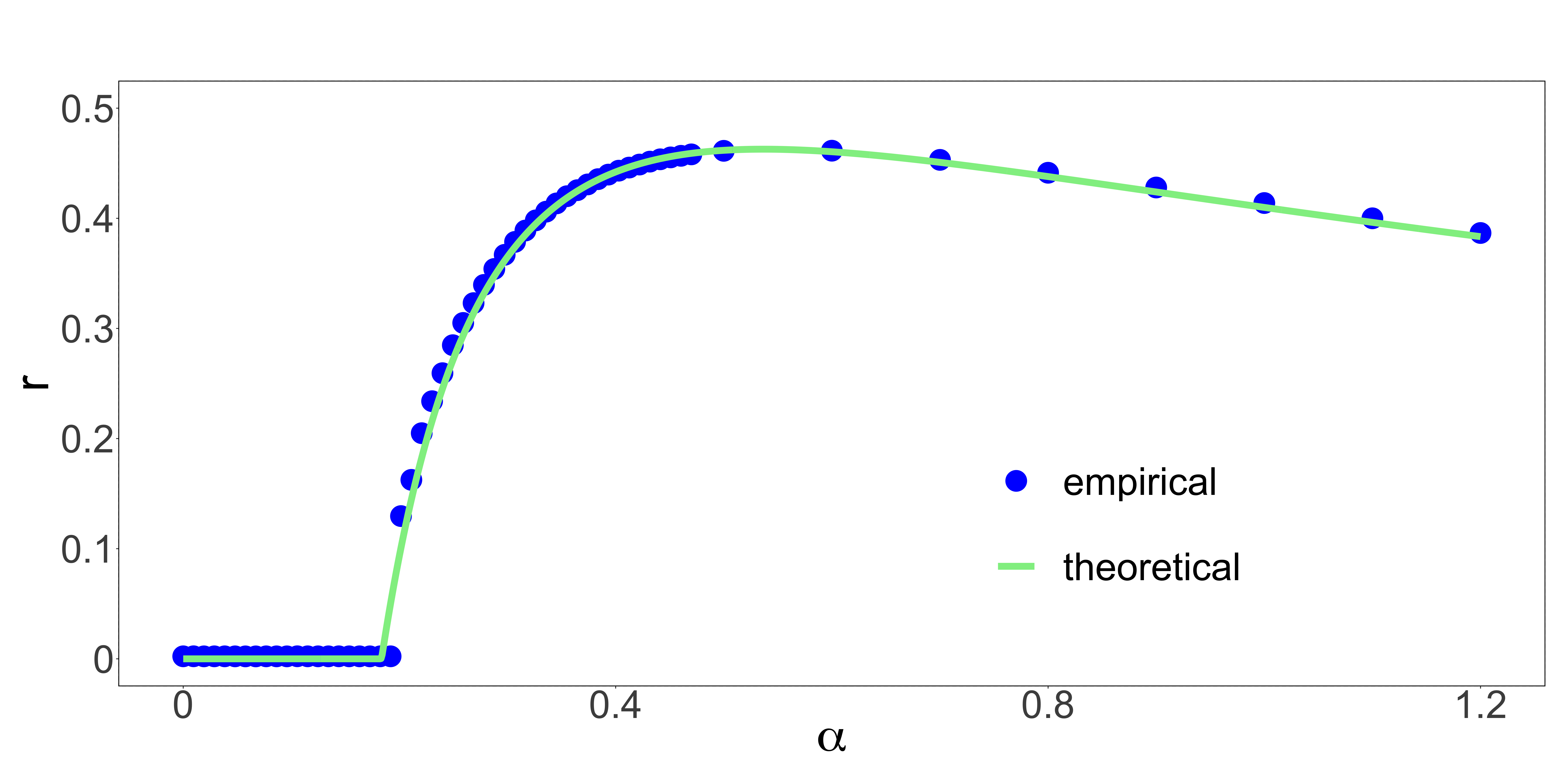}
	\caption{\label{fig:3} {\bf Emergence of cluster synchrony predicted by a self-consistent theory.} The heterogeneous mean-field approach and self-consistent theory lead to the definition of $R_{\alpha}(r)$ in Eq. (\ref{eq:R(r)}). Its bifurcation into a nonzero attractive fixed point $r>0$ pinpoints the emergence of cluster synchrony. The dots are empirically calculated values of the order parameter. The solid line shows the theoretic prediction of $r$ as the attractive fixed point of the self-consistent map $R_{\alpha}$, predicting the values of $r$ and the group of nodes in the cluster.}
\end{figure}

{\color{black} Our theory relies on the stationarity of ensemble amplitude $r$, which can break down for large coupling. The analysis of such cluster synchrony with non-stationary $r$ requires the development of a nonautonomous driving in the heterogeneous mean field approximation. If the network is small, {  finite-size effects in the heterogeneous mean field approximation can produce noise-induced phenomena. For instance, in a homogeneous network, where Ott-Antonson ansatz applies, it was found that finite-size effects can induce synchronization \cite{pikovsky2021} or delay synchronization \cite{Mendonca2023}.}}

{\it  In conclusion}, we have observed the spontaneous emergence of cluster synchronization towards an enslaving chaotic motion in a general class of systems, where the network is heterogeneous, the isolated maps chaotic, and the coupling function diffusive. {\color{black} In contrast to previous studies on cluster synchronization where an increasing number of nodes join the cluster for strong coupling, in our case of chaotic cluster synchronization, as the coupling increases, new nodes can join the cluster while certain nodes leave. We developed a heterogeneous mean-field approximation of the network effect on each connectivity layer and a self-consistent theory for the ensemble mean-field amplitude $r$. Our theory explains the emergence of cluster synchrony at the bifurcation of $r$ from zero into positive values. 
The prediction from our analysis is in excellent agreement with the empirically simulated $r $ from network dynamics. Our results could lead to a deeper understanding of collective dynamics in real-world networks with a heterogeneous topology that lacks symmetry.}

We thank  Edmilson Roque, Jeroen Lamb, Narcicegi Kiran, Thomas Peron, Serhiy Yanchuk for enlightening discussions. ZB and TP acknowledge support by FAPESP grants 2018/26107-0 and 2013/07375-0, Serrapilheira Institute (Grant No.Serra-1709-16124) and Newton Advanced Fellow of the Royal Society NAF$\backslash$R1$\backslash$180236). TP thanks the Humboldt Foundation via the Bessel Fellowship.

\appendix
\section{General chaotic cluster synchronization in heterogeneous networks} \label{sec_app:generalized}
We generalize the network dynamics into
\begin{equation}\label{eq:gen_network_dyn}
	z_p^{t+1}= f(z_p^t) + \frac{\alpha}{C}\sum_{q=1}^N A_{pq} \phi(z_q^t-z_p^t),
\end{equation}
where the node dynamics $f$ is chaotic and preserves the Lebesgue measure on [0,1). We can also analyze maps that preserve other measures but in this case the order parameter needs to be adapted; otherwise, we would obtain nonzero values of the order parameters even when the maps are uncoupled. The network $G$ has a heterogeneous degree distribution, and the coupling is diffusive in the sense that $\phi(0)=0$ and $\mathrm{D}\phi(0)\neq 0$.
In this case, it is known that for large networks {  \cite{Matteo}}  global synchrony is unstable for any $\alpha\neq 0$, as long as there is a nontrivial heterogeneity in the network $G$. 

We turn to cluster synchrony and use the ansatz that \textit{there is a sustained cluster $S_{\theta}$ synchronizing to the global field angle $\theta$ at coupling $\alpha$, while the other nodes spread uniformly.} 

\subsection{Heterogeneous Mean Field}
To compute the ensemble order parameter amplitude $ {r}$, we split it into contributions from different connectivity layers:
$$ {r}=\frac{1}{CN} \sum_{\mathrm{layers}~w}w\cdot \sum_{\mathrm{nodes}~q: ~w_q=w} e^{i2\pi (z_q- {\theta})}.$$

In the thermodynamic limit, (i) there are infinitely many layers $w$ following the degree distribution pdf $\delta(w)$, (ii) each layer $w$ has infinitely many nodes that distribute themselves on the circle according to some measure $\mu_w$.  Then, we have
\begin{align*}
	{r}= &\frac{1}{C} \sum_{\mathrm{layers}~w}  w \cdot \frac{1}{N} \sum_{\mathrm{nodes}~q:~w_q=w}e^{i2\pi (z_q-\theta)}\\
	\approx& \frac{1}{C}\int_0^{\infty}  w \delta(w)\cdot\int_{[0,1)} \mathrm{d}\mu_w(z) e^{i 2\pi (z-\theta)}.
\end{align*}

Generally the measure $\mu_w$ can be wild, the corresponding layer may contribute a nontrivial value 
$$\rho_w e^{i2\pi \psi_w}:= \int_{[0,1)} \mathrm{d}\mu_w(z) e^{i 2\pi z},$$ 
and these contributions may aggregate to a complicated $\theta$-dependent expression $$
r= \frac{1}{C}\int_0^{\infty} w\delta(w) \rho_w e^{i2\pi (\psi_w-\theta)}
$$

However, under the sustained cluster synchrony ansatz, the layers $w$ that lie outside the cluster $S_{\theta}$ spread uniformly so that $\mu_w=\mathrm{Leb}$ for most layers and hence
$$\int_{[0,1)} \mathrm{dLeb}(z) e^{i2\pi(z-\theta)}=0,~~~~\text{for any }\theta.$$
{\color{black} The heuristics of this reasoning is that the hub layer contains only a few nodes and does not contribute much to the order parameter. On the other hand, the cluster is formed around nodes in the connectivity layers near $C$. The layers with connectivity degree less than $C$ have $w_q/C\ll1$ and thus behave almost independently of the network. In fact, since the isolated dynamics is stochastically stable, they will distribute almost like Lebesgue}.
In other words, the non-cluster layers do not contribute to the ensemble order parameter amplitude $r$. On the other hand, the layers $w$ that lie inside the cluster $S_{\theta}$ synchronize to $\theta$ so that $\mu_w=\delta_{\theta}$ and hence
\begin{eqnarray}
	r&\approx& \frac{1}{C} \int_{\text{layers $w$ in }S_{\theta}} w\delta(w) \int_{[0,1)} \mathrm{d}\delta_{\theta}(z) e^{i2\pi (z-\theta)} \\
	&=& \frac{1}{C} \int_{\text{layers $w$ in }S_{\theta}} w\delta(w) \\
	&=&\frac{\mathrm{Vol}(S_{\theta})}{\mathrm{Vol}(G)}
\end{eqnarray}

Recall for node $p$  the local field is defined to be 
\begin{eqnarray}
	r_p e^{i2\pi \theta_p} &:=& \sum_{q=1}^N A_{pq} e^{i2\pi z_q} \\
	&\approx& \sum_{q\in S_{\theta}} A_{pq} e^{i2\pi z_q}\\
	&=& N(p,  S_{\theta}) e^{i2\pi\theta},
\end{eqnarray}
where $N(p,  S_{\theta})$ is the number of neighbors of node $p$ that belong to cluster $ S_{\theta}$. By concentration of the random graph, we may approximate this number by the ensemble average
\begin{align*}
	N(p,  S_{\theta}) \approx &\mathbb{E}[N(p,  S_{\theta})] = \sum_{q\in S_{\theta}}  \mathbb{E} A_{pq} \\ 
	=& \sum_{q\in  S_{\theta}} \frac{w_p w_q}{CN} = w_p \frac{\mathrm{Vol}( S_{\theta})}{CN} \\
	=& w_p r.
\end{align*}
We have  deduced
$$r_p e^{i2\pi \theta_p} = w_p r e^{i2\pi\theta}.$$
Similarly we have 
$$\sum_{q\in S_{\theta}} A_{pq} e^{i2\pi k z_q} = w_p r e^{i2\pi k\theta},~~~~p\in\{1,\cdots,N\},~~k\in\mathbb{Z}.$$
With Fourier expansion $\phi(x)=\sum_{k\in\mathbb{Z}} a_k e^{i2\pi kx}$, we compute
\begin{align*}
	z_p^{t+1} = & f(z_p^t) + \frac{\alpha}{C}\sum_{q=1}^NA_{pq} \phi(z_q^t-z_p^t)\\
	=& f(z_p^t) + \frac{\alpha}{C} \sum_{k\in\mathbb{Z}}a_k e^{-i2\pi k z_p^t} \sum_{q=1}^NA_{pq} e^{i2\pi k z_q^t}\\
	=& f(z_p^t) + \frac{\alpha}{C} \sum_{k\in\mathbb{Z}}a_k e^{-i2\pi k z_p^t} \sum_{q\in  S_{\theta}}A_{pq} e^{i2\pi k \theta^t}\\
	=& f(z_p^t) + \frac{\alpha}{C} \sum_{k\in\mathbb{Z}}a_k e^{-i2\pi k z_p^t} w_p r e^{i2\pi k \theta^t}\\
	=& f(z_p^t) + \frac{\alpha}{C} w_p r  \phi(\theta^t-z_p^t).
\end{align*}

\subsection{Master Stability Function}

To determine stability of the cluster $S_{\theta}$, we analyze the dynamics of small perturbations about the synchronous motion $s=z-\theta$ with respect to the global field angle. Consider
\begin{align*}
	s_p^{t+1} = z_p^{t+1} - \theta^{t+1}= f(z_p^t) + \frac{\alpha}{C}w_p r \phi(\theta^t-z_p^t) - {f}(\theta^t).
\end{align*}
Generally we cannot find a synchrony map like $f_{\beta}$ which evolves $s^{t+1}=f_{\beta}(s^t)$ as in the linear case $f(x)=2x\mod1$. We linearize at $s_p=0$ to obtain
\begin{align*}
	s_p^{t+1}=(\mathrm{D}f(\theta^t) - \beta_p \mathrm{D}\phi(0))s_p,~~~~\beta_p:=\frac{\alpha}{C} w_p r,
\end{align*}
The stability of $s_p=0$ translates to the stability of the cluster.  To determine the stability of the trivial solution, we consider the so-called Master Stability Function ($\mathrm{MSF}$) mapping effective coupling strength $\beta$ to the largest Lyapunov exponent of the multiplier cocycle $\mathrm{D}f(\theta^t) - \beta_p \mathrm{D}\phi(0)$. We have thus identified the synchrony condition for node $p$ to be 
$$p\in S_{\theta}\iff\mathrm{MSF}\left(\frac{\alpha}{C}w_p r \right)<0.$$

\subsection{Generalized self-consistent equation}
Similar to the linear case $f(x)=2x\mod1$, the self-consistent equation for $r$ reads
$$R_{\alpha}(r)= \frac{1}{C}\int_{\{w: \mathrm{MSF}(\alpha w r/C)<0\}} w \delta(w)\mathrm{d}w,$$
where $\delta(w)$ is the degree distribution density function.

{ \subsection{Further examples: tent map}\label{sec_app:tent}}
As an illustration, consider the tent map as node dynamics
$$f(x):=\begin{cases}
	2x,& x\in[0,1/2];\\
	2(1-x),& x\in[1/2,1],
\end{cases}$$
and $\phi(x):=\sin(x)$ coupled on a random graph on $N= 5\times 10^4$ nodes with inverse gamma $\mathrm{Inv}\Gamma(2,C)$ degree distribution. This gives
\begin{align*}
	\mathrm{D}f(\theta^t) - \beta \mathrm{D}\phi(0) =& 2(\boldsymbol{1}_{(0,1/2)}- \boldsymbol{1}_{(1/2,1)})(\theta^t) - \beta,
\end{align*}
where $\boldsymbol{1}_I$ is the indicator function of the interval $I$.

It can be observed that the interval $(\beta^-,\beta^+)=(1/2\pi,3/2\pi)$ is the stability region for effective coupling strengths. We identify the synchronous layers
$$w= \frac{\beta C}{\alpha r } \in\left(\frac{\beta^- C}{\alpha r } ,\frac{\beta^+ C}{\alpha r } \right).$$
Thus, we obtain the self-consistent equation for cluster synchronization in network system with tent map as node dynamics:
\begin{align*}
	R_{\alpha}(r)=& \frac{1}{C} \int_{\beta^- C/ \alpha r}^{\beta^+ C/\alpha r} w \cdot C^2 w^{-3} e^{-C/w}\mathrm{d}w\\
	=& e^{-\alpha r/\beta^+} - e^{-\alpha r/\beta^-}.
\end{align*}

\section{Synchrony map} \label{sec_app:sync_map}
Recall that the synchrony map 
\[f_{\beta}(s)=2s-\beta\sin(2\pi s)\mod1
\] 
governs the evolution of the relative state $s=z-\theta$ with respect to the global mean field angle $\theta$, where the effective coupling strength $\beta=\frac{\alpha}{C} w r $ depends on the network coupling strength $\alpha$, mean degree $C$, degree $w$ of the node and the stationary global field amplitude $r $ at $\alpha$.  By differentiating $f_{\beta}$ with respect to $s$, we have
$$\left.\frac{\mathrm{d}f_{\beta}}{\mathrm{d}s}\right|_{s=0}=2 - \beta 2\pi   \cos(  2\pi s)|_{s=0}=2- \beta 2\pi.$$

This gives the lower bound $\beta_1=1/2\pi$ and upper bound $\beta_2=3/2\pi$ for the stability region, that is, the fixed point $s=0$ attracts when the effective coupling strength is tuned to be $\beta\in (\beta_1,\beta_2)$, and the attraction is strongest with derivative zero when the effective coupling strength is $\beta_0=1/\pi$.

{ Through a more detailed bifurcation analysis,} four parameter regimes can be observed:
\begin{enumerate}
	\item for small $\beta\in[0,\beta_1]$, the fixed point $s=0$ is not attractive. The corresponding layers $w=\frac{\beta C}{\alpha r }\in[0, \frac{\beta_1 C}{\alpha r}]$ are not enslaved to the motion of global field angle $\theta$;
	\item for  $\beta\in (\beta_1,\beta_2)$, the fixed point $s=0$ attracts exponentially. Provided that the stationary $r>0$, the corresponding layers $w=\frac{\beta C}{\alpha r}\in (\frac{\beta_1 C}{\alpha r}, \frac{\beta_2 C}{\alpha r})$ synchronize by virtue of enslavement towards the global field angle $\theta$;
	\item for larger $\beta\in[\beta_2,\beta_3]$ where $\beta_3\approx 0.6$, the fixed point $s=0$ loses stability and gives rise to an attractive period-2 orbit around it. The corresponding layers $w=\frac{\beta C}{\alpha r}\in (\frac{\beta_2 C}{\alpha r}, \frac{\beta_3 C}{\alpha r})$ are enslaved towards a neighborhood of the global field angle $\theta$, jumping around it in a period-2 fashion. This explains the funnel shape of layers above the synchronized cluster in Fig. 1.
	\item for large $\beta>\beta_3$, the period-2 orbit undergoes a cascade of period doubling bifurcations to enter a chaotic regime with windows of stability. In a finite network, the corresponding highly connected layers $w=\frac{\beta C}{\alpha r}>\frac{\beta_3 C}{\alpha r}$ feel little effect of the synchrony map due either to the chaotic regime or to finite-size effect.
\end{enumerate}

It is important to have a \textit{positive} and \textit{stationary} global field amplitude $r$ in order to pass from the bifurcation analysis of $f_{\beta}$ to the corresponding layers in the network dynamics defined with $r$ in the denominator. Indeed, $r=0$ suggests that the network coupling strength $\alpha$ in question does not support cluster synchrony. And when the global field amplitude is non-stationary, see the {  Appendix \ref{sec_app:self-consistent}}.

For network coupling strengths $\alpha$ that admit cluster synchrony, that is, when stationary amplitude $r>0$, the node dynamics are enslaved to the doubling motion $\theta\mapsto f(\theta)$ of the global mean-field angle. More precisely, consider the skew-product for the layer with degree $w$
$$F(\theta,z)=(f(\theta), f(\theta)+f_{\beta}(z-\theta)),~~~~\beta:=\frac{\alpha}{C}wr.$$
A moderate layer $w\in(\beta_1 C/\alpha r,\beta_2 C/\alpha r)$ enjoys effective coupling strength $\beta\in (\beta_1,\beta_2)$, for which the synchrony map $f_{\beta}$ shrinks the relative distance $(z-\theta)$ to 0 exponentially fast. At a uniformly distributed initial network state, the initial global mean-field magnitude $r^0\ll 1$ is small and hence the synchrony maps first brings highly connected nodes with degree $w=C/2\pi \alpha r^0\gg 1$ into coherence towards $\theta$. As $r^t$ increases in the meantime, the synchrony map loses control over these highly connected layers and moves to enslave lower layers. At stationarity $r$, the synchrony map fully captures the synchronized cluster and sustains the cluster synchrony configuration. Fig. 2 shows the time-evolution of the cluster being captured and thereby entering sustained synchrony.

\section{Self-consistent theory}\label{sec_app:self-consistent}
Recall the synchrony cluster
{ $$S_{\theta}=\{i:w_p\in (C \beta^-/\alpha r,C \beta^+/ \alpha r )\},$$
	with $(\beta^-,\beta^+)=(1/2\pi,3/2\pi)$}
which leads to the definition of the self-consistent equation
\begin{align*}
	R_{\alpha} (r)
	=& \int_{C \beta^-/ \alpha r}^{C\beta^+/\alpha r} C w^{-2} e^{-C/w}\mathrm{d}w,~~~~\delta(w)=C^2 w^{-3} e^{-C/w}\\
	=& e^{-\alpha r/\beta^+} - e^{-\alpha r/\beta^-}.
\end{align*}

{ 
\begin{enumerate}
\item[(i)] At low connectivity level, the nodes are close to being uniformly distributed. Their contribution to the mean-field amplitude $r$ is negligible.

\item[(ii)] Near but below the cluster layers, the mean-field skews the layer distributions towards zero, thus contributing to the mean-field amplitude $r$. 

\item[(iii)] The cluster with effective coupling strength $[\beta^-,\beta^+]$  synchronize. These are the layers considered in Eq. (\ref{eq:R(r)}). 

\item[(iv)] Near but above the cluster layers, the synchrony map undergoes a period doubling cascade, causing these layers also to contribute to the mean-field amplitude $r$. Our hand tuning accounts for them by including effective coupling strengths as high as $\beta^+ +0.049$.

\item[(v)] At high connectivity levels, there are few such massive hubs and their contribution to $r$ is also negligible. 
\end{enumerate}
}

To generate the solid curves in Fig. 3, we use successive approximation: for each $\alpha$ probed in $[0,1.2]$ spaced $0.001$ apart, we initialize at $r^0=0.1$, iterate $1000$ times by $R_{\alpha}$ and output the last value as  $r $.
{  In the derivation of $R_{\alpha}$, the only occasion using the inverse gamma degree distribution $\delta(w)$ is at the second step. For other heterogeneous degree distributions, such as for Barab\'asi-Albert network, the same derivation can be performed, with $\delta(w)$ replaced accordingly. In a homogeneous network, such as a small-world in the sense of Watts-Strogatz, this calculation is not so meaningful, as nodes have the same connectivity. We expect full synchrony in this case.} 


\section{Forcing a large network into synchrony}\label{sec_app:forced_sync}
Consider a large network, in our case, $N=5\times 10^5$. Even at a coupling strength $\alpha$ that admits cluster synchrony, i.e., $r >0$, it still may take a prohibitively long time for the network to evolve spontaneously from a uniformly distributed initial state into cluster synchrony. To deal with this issue in our numerical experiments, we prepare the initial network condition at a certain synchrony level $r^0\in (0,r )$ by pointing a suitable cluster of nodes all toward phase 0. Such a prepared initial low level of synchrony serves to spark the cluster synchrony, which, once in motion, is allowed to run freely. 

\bibliography{references}
\end{document}